\documentclass{article}
\usepackage{sao1}
\usepackage{graphicx}
\usepackage{longtable}
\begin{document}

\title{Properties of a Volume-Limited Sample of Magnetic Ap/Bp Stars}
\author{Power, J.\inst{1,2} \and Wade, G. A.\inst{1} \and Hanes, D. A.\inst{2} \and 
  Auriere, M.\inst{3} \and  Silvester, J.\inst{1,2}}
\institute{ Royal Military Collage of Canada
	    \and Queen's University
	    \and Observatoire Midi-Pyrenees }
\maketitle

\begin{abstract}
This paper describes a study to deduce fundamental parameters and magnetic field characteristics for all magnetic Ap/Bp stars with in a 100 parsec heliocentric radius volume.  This study has allowed for the first time the determination of an effectively unbiased magnetic field distribution of a sample of intermediate mass stars.  From published catalogues and other literature sources, we have identified 57 {\em bona fide} magnetic A and B stars in the volume, corresponding to 1.7\% of all intermediate mass stars within 100 parsec of the Sun.  The masses of Ap stars range from 1.5 to 6~$M_\odot$, with the peak of the mass incidence distribution between 3.3 and 3.6 $M_\odot$.  Observations of 30 of the Ap/Bp stars were obtained using the MuSiCoS spectropolarimeter at the Telescope Bernard Lyot at Pic du Midi Observatory.  These observations will be used to refine periods, and determine magnetic field strength and geometry.
\end{abstract}

\section{Introduction}

Ap/Bp stars are a class of intermediate mass stars which display magnetic fields and characteristic chemical peculiarities.  These stars show overabundances typically of iron peak elements, rare earths, and silicon, ranging up to $\sim$2 dex above solar.  These magnetic chemically peculiar stars make up a few percent of the main-sequence A and B stars.

  Ap/Bp have globally ordered, predominantly dipolar magnetic fields, with strengths of the order of one kilogauss.  The presence of strong, ordered magnetic fields in some main-sequence A and B stars has been known for nearly one-half of a century (Babcock 1947).  However the cause of this magnetic field still remains a mystery.  Intermediate mass stars are not expected to display magnetic fields, as they lack the deep convective envelopes which drive the dynamo effect responsible for magnetic fields seen in lower mass stars like the sun.  

  There are two competing theories for the generation of the globally ordered magnetic field seen in Ap/Bp stars: the contemporaneous dynamo effect, and the fossil field field theory.  Contemporaneous dynamo effect theorizes that there is a dynamo effect currently working in the (convective) core of the star.  The fossil field theory hypothesizes that the magnetic field is a remnant, produced by a dynamo effect operating at an earlier evolutionary phase, or swept up from the interstellar medium during star formation. 

The aim of this study to provide an essentially unbiased assay of the properties of magnetic A and B stars in the solar neighbourhood.  By deducing fundamental parameters and magnetic field characteristics for all magnetic intermediate mass stars within 100 pc heliocentric radius volume, this study will allow or the first time the determination of effectively unbiased statistical study of a sample of Ap/Bp stars.

\section{Sample stars and properties}

\subsection{Magnetic A and B Stars Selection}

With the availability of accurate parallaxes for nearby stars through the Hipparcos catalogue (ESA 1997), a distance limited sample was determined. A catalogue of all stars within 100 parsecs of the sun was compiled from the Hipparcos database.  This list was then filtered for stars with B-V colour index between 1.0 and -0.4, which corresponds to main sequence stars ranging from approximately O to K spectral types (Gray 2005).  The broad range in colour index was chosen to ensure that all intermediate mass stars in the volume were included, while limiting the size of the database which we are required to analyse.

From this distance limited sample, Ap/Bp stars were selected through comparison with Renson's {\it General Catalogue of of Ap and Am stars} (Renson et al. 1991).  Only magnetic chemically peculiar stars were considered for this study, corresponding to Renson peculiarity types SrCrEu, Si, He weak Si, SrTi, and He strong (the types of chemically peculiar stars known to display magnetic fields).  This requires some care, as the peculiarity type of some stars is incorrectly indicated (e.g. HD 117025 is quite clearly an Ap star, but appears as Am in Renson's list). Using Renson's catalogue, cross-referencing with the distance limited sample, and verifying the Ap/Bp nature of each candidate in the literature, 52 {\em bona fide} Ap/Bp stars were found to be within 100 pc of the sun. Essentially all of these stars are bright and well-studied.  
  
To ensure that no magnetic A and B stars were overlooked, a cross-check with the {\it Bright Star Catalogue} (Hoffleit et al. 1991) was performed.  All A, B, and late F stars in the {\it Bright Star Catalogue} with a peculiarity note were examined in the literature for photometric variability and/or magnetic field detection.  Five stars not listed as Ap/Bp stars in Renson's catalogue were added to the sample.  The final sample of magnetic Ap/Bp stars within 100 pc heliocentric radius volume was found to consist of 57 stars. This sample is summarised in Table 1.

\subsection{Physical parameters of sample stars}

Stromgren and/or Johnson photometry were obtained for each of the magnetic and non-magnetic stars in the complete distance limited sample from the {\it General Catalogue of Photometric Data} (Mermilliod 1997).  Temperatures and luminosities were determined for the non-magnetic stars using various photometric calibrations.  The Stromgren-based photometric calibration of Balona (1994) for spectral types O through F was used to determine temperatures and bolometric corrections when $uvby\beta$ photometry was available. A suitable and reliable Geneva calibration which covered a large temperature range was never found for the non magnetic sample.  However, the majority of stars which had Geneva photometry also had Stromgren photometry, for which Balona's (1994) calibration was used.  The Johnson B-V calibration from Gray (2005) was also used for the non-magnetic for those stars of which Stromgren photometry was unavailable.

For the Ap/Bp star sample, temperatures were determined using a variety of techniques.  For some stars, observed relative energy distributions (Adelman et al. 1989) were fit using a grid of model ATLAS9 flux profiles. For those stars for which observed energy distributions were unavailable, several Stromgren and Geneva photometric calibrations were used and averaged to determine the temperatures (Balona 1994, Stepien 1994, Hauck and North 1993, Kupka and Bruntt 2001).  The average uncertainty was found to be just under 500 K, consistent with the "realistic" temperature determination uncertainty for Ap stars suggested by Landstreet (this volume).  Luminosites was determined using a temperature-based BC calibration:
\begin{equation}
	BC_{Ap} (\theta ) = -4.891 + 15.147 \theta - 11.517 \theta^2 
\end{equation}
\noindent where $\theta$=5040.0/$T_{\rm e}$ (Landstreet 2006).  This bolometric correction best describes Ap stars between 7500K and 18000K.  The uncertainty associated with using this calibration is about 0.1 mag (Landstreet 2006).

Finally, a literature review was conducted to compile a database of other physical parameters of the Ap/Bp sample, such as projected rotational velocity (Royer et al. 2002, Abt et al. 2002, Brown and Verschueren 1997, Abt and Morrell 1995, Bohlender et al 1993), magnetic field strength (Bychkov et al. 2005, 2003, Romanyuk 2000, Shorlin et al 2002), and rotational period (Catalano and Renson 1998, Bychkov et al. 2005, ESA 1997).

\subsection{HR Diagram}    
\begin{figure}[!htb]
\centering
\includegraphics[width=11.8 cm]{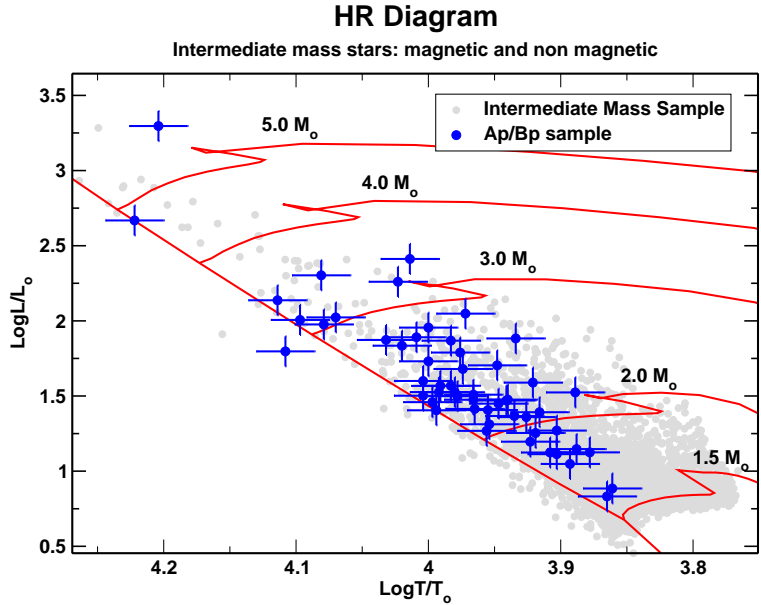}
\caption{\it Points with error bars represent the distance limited sample of magnetic A \& B stars.  The gray points shows the position of the non-magnetic intermediate mass sample, with a lower cut off at 1.5 $M_\odot$. The lower cut off is chosen based on the lowest mass of the Ap/Bp stars in the sample. The solid lines show the theoretical evolutionary tracks and ZAMS (Schaller et al. 1992)}
\label{HRDiagram}
\end{figure}

The magnetic and non magnetic star samples were plotted on a Herzsprung-Russell diagram (Figure \ref{HRDiagram}).
 
In Figure 1, the star positions on the HR diagram, with their uncertainties, are compared with the evolutionary tracks of Schaller et al. (1992) for Z=0.02.  Theoretical evolutionary tracks were interpolated between those provided by Schaller et al. for completeness.  The masses of the magnetic A and B sample were determined by manually comparing their position on the HR diagram with that of the tracks, while the masses of the larger sample were computed through an automated process which assumed mean uncertainties for effective temperature and luminosity.

  Masses were determined for 54 of the 57 stars in the magnetic A and B sample.  HD 12709 and HD 133652 lie below the calculated models on the HR diagram, so additional study of these stars is required.  No available photometry was found for HD 107452, so no temperature was determined, and hence no mass is available for this star.  This star will also be examined in further detail.

  The larger sample of non-magnetic stars was filtered to focus on the mass range of interest with a lower limit of 1.5 $M_\odot$, chosen based on the lowest-mass stars in the magnetic A \& B star sample. We find that there exists a significant number of magnetic Ap/Bp stars (2 stars) in the sample in the mass range of 1.5-1.8 $M_\odot$, making 1.5 $M_\odot$ in reasonable lower limit for the intermediate mass sample. The final distance limited intermediate-mass sample includes 3297 stars, with masses ranging between 1.5 $M_\odot$ and 10 $M_\odot$.  

\section{Statistics}

\subsection{Period Distribution}

\begin{figure}[!htb]
\centering
\includegraphics[width=11.8 cm]{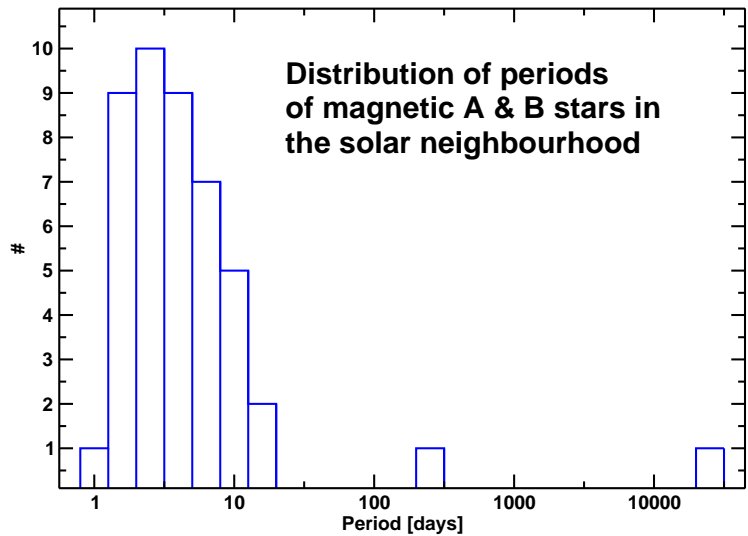}
\caption{\it The period distribution for the magnetic Ap \& Bp stars in the solar neighbourhood.  The period is given logarithmically across the horizontal axis, and the number in each bin is labeled across the vertical axis.}
\label{periods}
\end{figure}

  Periods were found in the literature for 49 of the 57 magnetic Ap and Bp stars in the solar neighbourhood. The period distribution of the Ap/Bp sample revealed a strong tendancy to rotation periods of order 1 day, with few exceptions (Figure \ref{periods}).  The mean period was found to be 4.5 days, neglecting the 2 stars with significantly longer periods, with a standard deviation of 4.0 days. The distribution displays a clear peak between 3.3 to 5.0 days with steep drop off at shorter periods, and a more gradual drop off at longer periods.  Two stars had periods longer than 100 days.  The roAp star $\gamma$ Equ is a well known extreme slow rotator, with a period of 91.1 years (Bychkov et al. 2006).  HD 134214, another well-known roAp stars, has a period of 268 days (Kreidl et al. 1994)

\subsection{Mass Distribution}

\begin{figure}[!htb]
\centering
\includegraphics[width=7.0 cm]{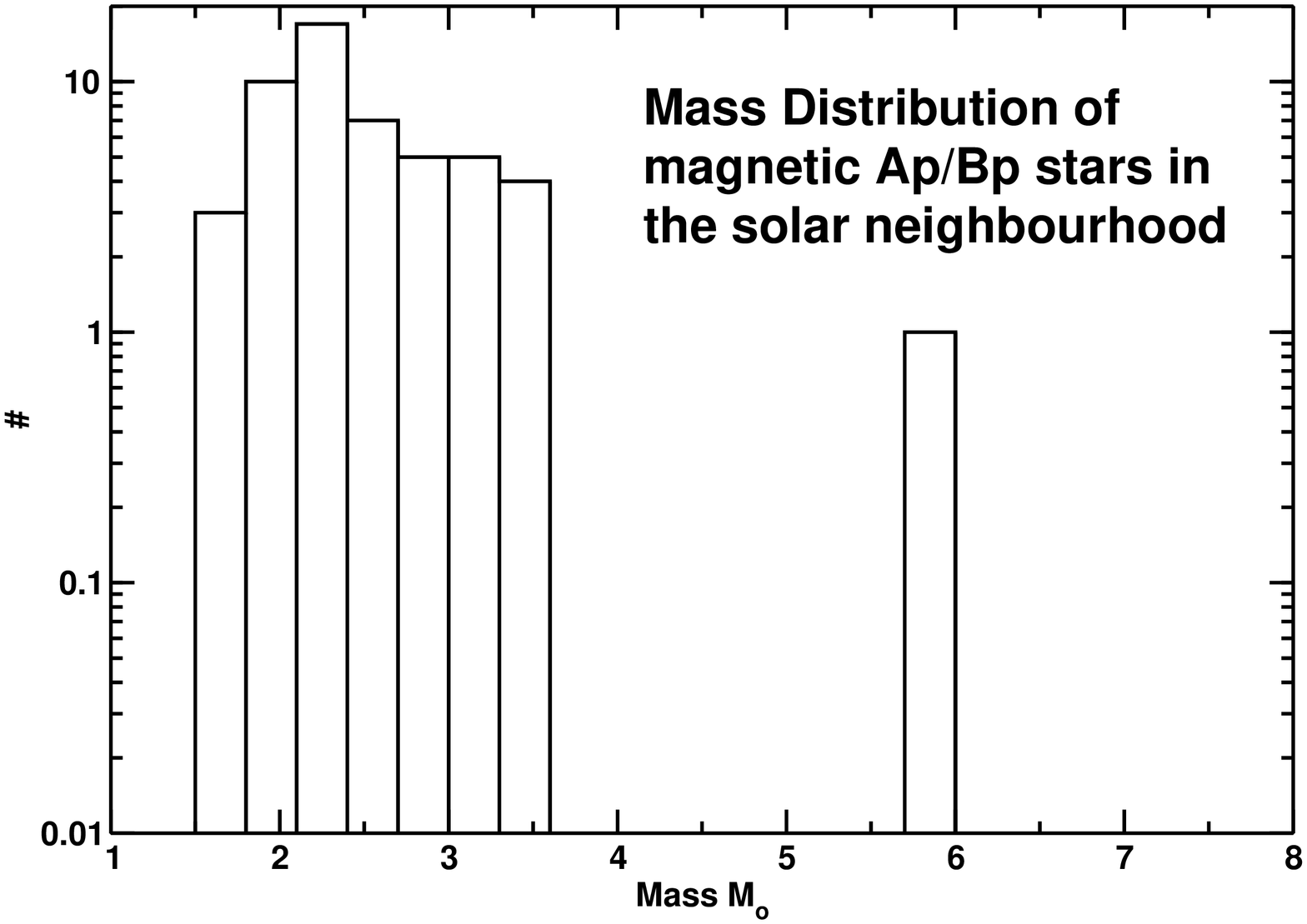}\hspace{1mm}\includegraphics[width= 7.0cm]{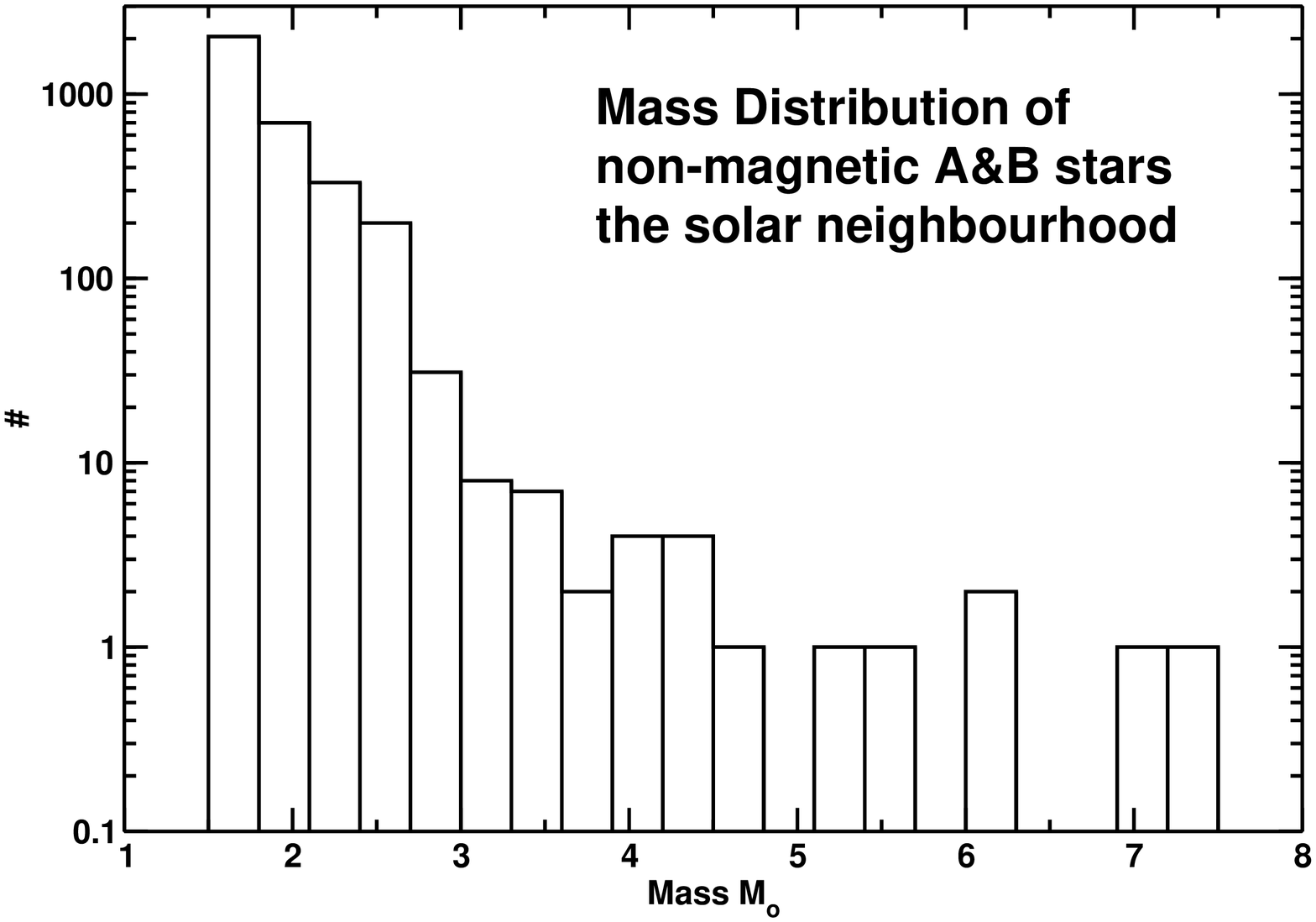}
\caption{\it Right: Mass distribution of magnetic A \& B stars within 100 pc heliocentric radius.  Left: Mass distribution of non-magnetic intermediate mass stars in the solar neighbourhood. }
\label{massdistribution}
\end{figure}

The mass distribution of Ap/Bp stars in the solar neighbourhood (Figure \ref{massdistribution}) shows a clear maximum between 2.1 and 2.4 $M_\odot$.  The lower limit of the Ap/Bp distribution is at 1.5 $M_\odot$.  The Ap/Bp mass distribution is clustered between 1.5 and 3.6 $M_\odot$ with one exception at 6.0 $M_\odot$.

By contrast, the number of non-magnetic intermediate mass stars with a particular decreases steeply with increasing mass.   

\begin{figure}[!htb]
\centering
\includegraphics[width=11.8 cm]{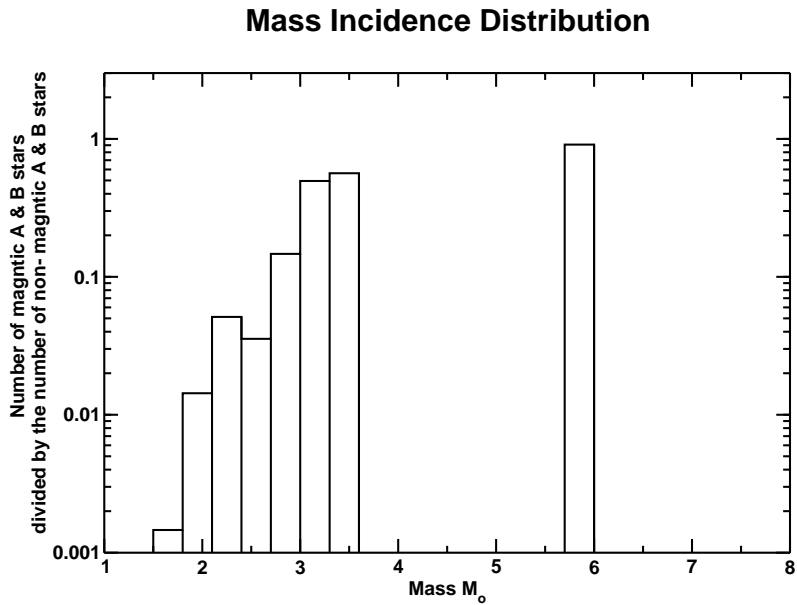}
\caption{\it  The mass incidence of Ap/Bp stars, determined from the mass distribution of the Ap/Bp sample and the intermediate mass sample.  Above 3.6 $M_\odot$, a sufficiently large sample of stars is not available to provide significant statistics.}
\label{massIncidencedistribution}
\end{figure}

The mass incidence distribution of Ap/Bp stars was calculated by dividing the mass distribution of Ap/Bp stars by that of the non-magnetic stars (Figure \ref{massIncidencedistribution}.  The mass incidence distribution shows a clear peak at 3.3 to 3.6 $M_\odot$. Above $3.6~M{\odot}$, the number of stars in the sample is sufficiently small that the shape of the mass incidence distribution is completely uncertain.  

  Of the 3297 stars with masses above the lower limit of 1.5 $M_\odot$ within 100 parsecs of the sun, 57 were found to be Ap/Bp stars. Hence the bulk incidence of of magnetic Ap/Bp stars in the solar neighbourhood is 1.7\%. This value is substantially smaller than the incidence reported by classical studies of magnitude-limited samples (Wolff 1968, Johnson 2004). We are currently attempting to estimate realistic uncertainties associated with this value. If the Ap/Bp sample and non-magnetic stars sample are both uncertain by 10\%, the uncertainty is 0.4\%.

\section{Observations and Analysis}

\begin{figure}[!htb]
\centering
\includegraphics[width=5.4 cm]{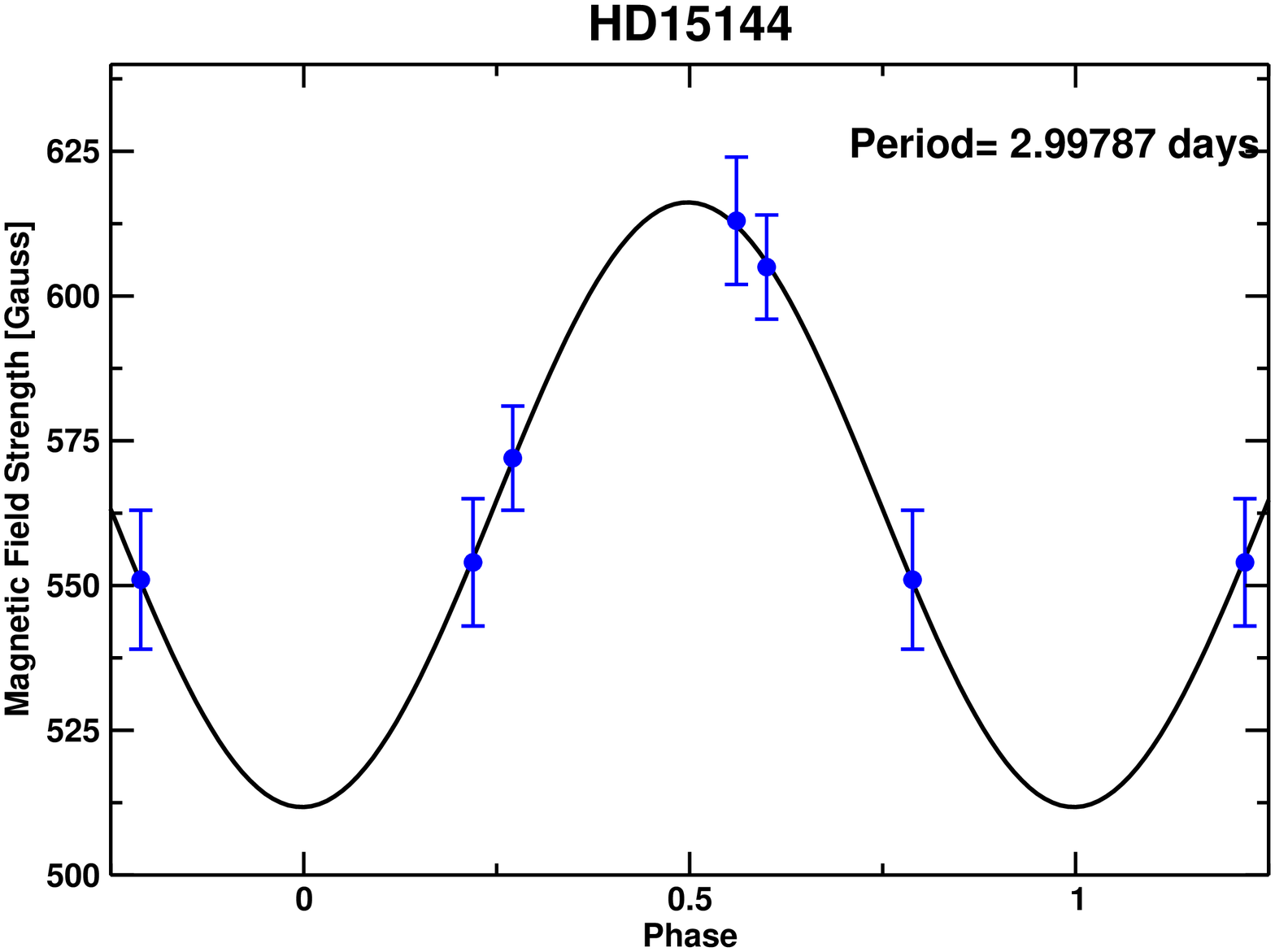}\hspace{1mm}\includegraphics[width=5.4 cm]{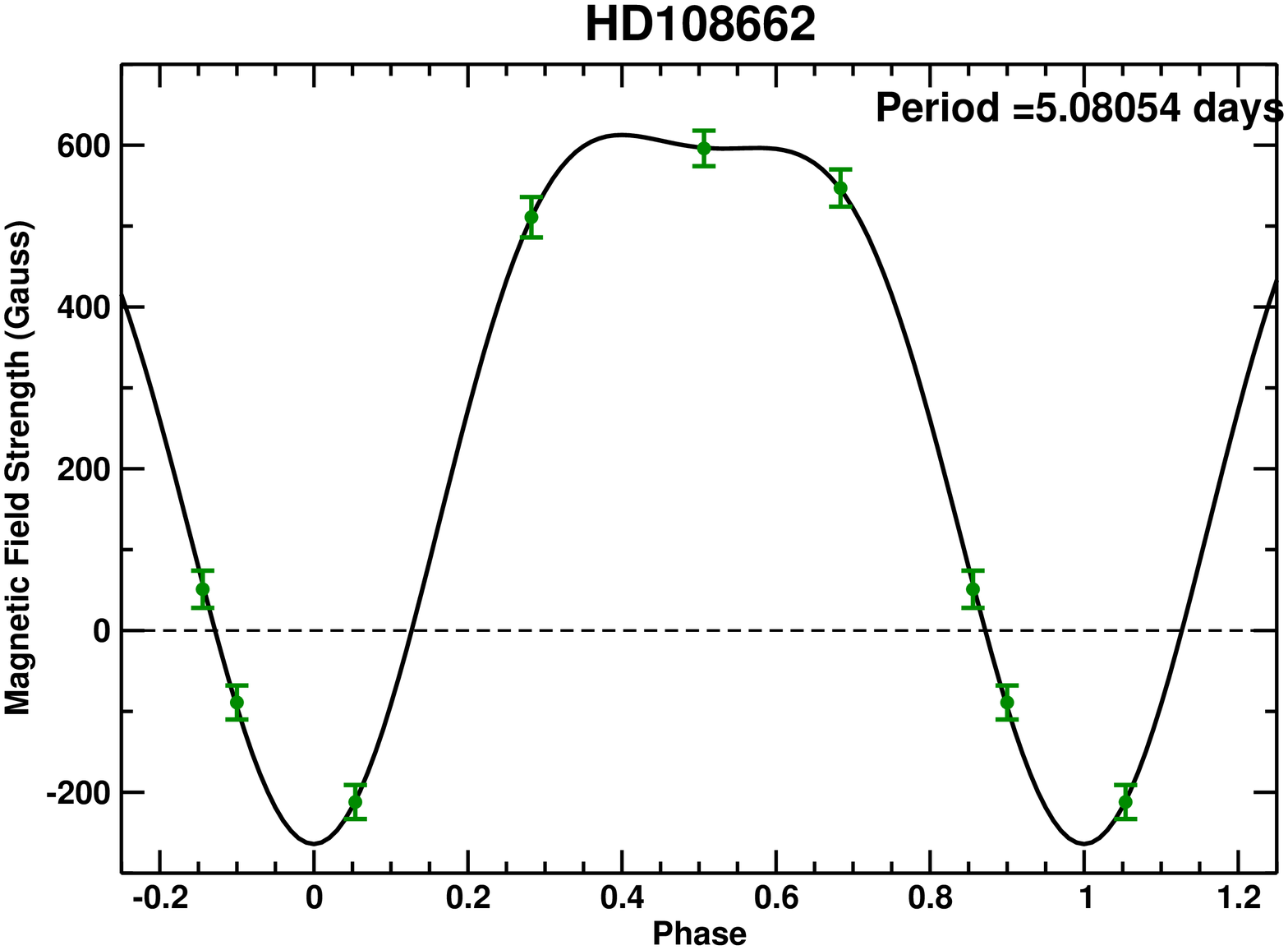}\hspace{1mm}\includegraphics[width=5.4 cm]{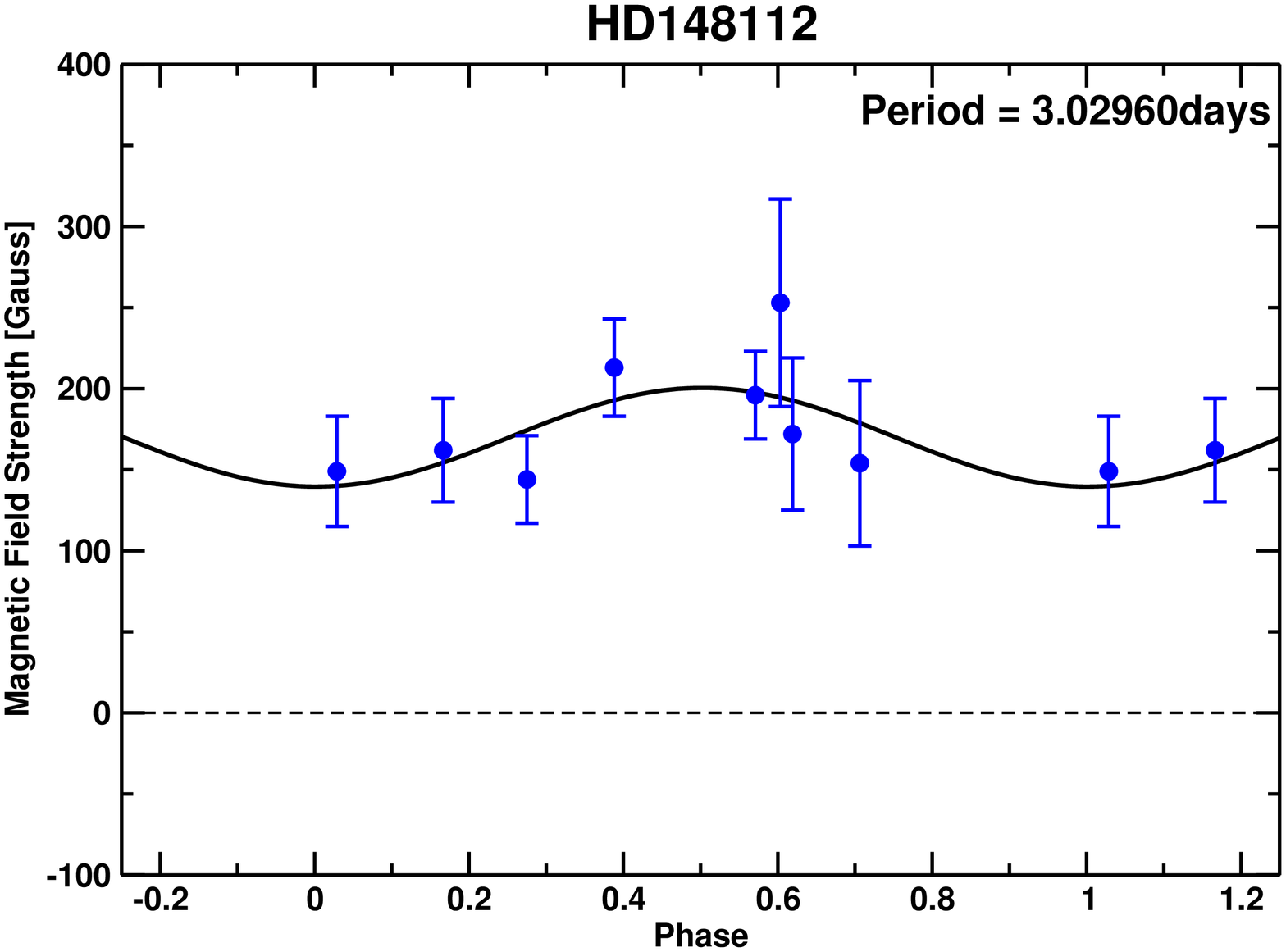}
\caption{\it Right:  HD15144 is a spectroscopic binary.  The period was determined photometrically by van Genderen (1971) to be 2.997814 days, which fits well with the observed variation of the longitudinal magnetic field determined from observations.  Middle: HD108662 shows strong magnetic variability of the longitudinal magnetic field.  The rotational period determined by Rice and Wehlau (1994) fit well with a second order sinusoid to the observations.  Left: HD148112 star exhibits weak magnetic variability and a nearly constant longitudinal field.  The period of 3.04296 days was determined spectroscopically by Hatzes (1991)}
\label{perioddetermination}
\end{figure}

Observations of the Ap/Bp sample were obtained using the MuSiCoS spectropolarimeter at the Telescope Bernard-Lyot at Pic du Midi Observatory.  Reduction of the circular polarization spectra was performed using the ESPRIT data reduction software package (Donati et al. 1997)  Least squares deconvolution (LSD), a multi-line analysis technique, developed by Donati et al. (1997), was used to extract the mean Stokes I and Stokes V profiles from each spectrum.  The mean longitudinal field for each of the observations was determined by taking the first moment of the Zeeman split Stokes V profile (Wade et al. 2000).  The longitudinal magnetic field was plotted as a function of phase (Figure \ref{perioddetermination}). 
 
\section{Future Work}

Using the observations obtained with TBL-MuSiCoS, parameters such as projected rotational velocity, period, and magnetic field strength and geometry will be derived for all 30 observed Ap/Bp stars.  Periods for each of the observed stars will be confirmed and refined.  The volume radius may be extended to 200 parsecs to improve the statistics at higher masses.  By extending the volume to 200 parsecs the volume sample would include the Sco-Cen association, and thus would include more higher mass Ap stars.  

\section{Conclusion}
Fundamental parameters have been determined for a complete sample of intermediate mass stars within 100 parsecs heliocentric radius.  The Ap/Bp stars have been carefully identified for this sample and an examination of the statistical properties of these stars has been performed.  Magnetic intermediate mass stars are found to make up 1.7\% of all intermediate mass stars in the solar neighbourhood.  These stars typically show magnetic and/or photometric variability on the order of several days, but in some cases variability periods are much longer.  For this particular sample, the masses of the magnetic A \& B stars ranged from 1.5 to 6 $M_\odot$, peaking between $3.3$ and $3.6~M_{\odot}$. New spectropolarimetric observations have been obtained for essentially all sample AP/Bp stars observable from the northern hemisphere. These data will be used to refine $v\sin i$ estimates, rotational periods, and to determine magnetic fields strengths and geometries. We are planning to extend the sample to 200 parsec heliocentric radius in order to refine the statistics at higher masses.

\newpage

\begin{center}
{\it Table 1: The distance limites sample of Ap/Bp stars.  The spectral types given are those listed by Renson et al. (1991) if available, otherwise they were obtained from the Hipparcos catalogue (ESA 1997).  The calculated temperature for each star is given. 1.) calculated from fitting model flux profiles to energy distributions of Adelman et al. 1989.  Averaged photometric calibrations were used otherwise: 2.) Kupka \& Bruntt (2001), 3.) Hauck \& North (1993), 4.) Balona (1994), 5.) Stepien (1993).  Derived masses are indicated. Periods, when available, are listed for each of the sample stars. Periods derived from literature: 6.) Bychkov et al. (2005), 7.) Catalano \& Renson (1998), 8.) ESA (1997).  }
\end{center}
\begin{longtable}{c|c|c|c|c|c}

HD & Spectral Type & Temperature (K) & log(L/$L_\odot$) & Mass ($M_\odot$) &Period (days) \\
\hline
\endhead

1185 & A1SI & 9012 $^{ 2,3,4 } $ & 1.406 & 2.15 &    \\
3980 & A7 SR EU CR & 8291 $^{ 2,3 } $ & 1.254 & 1.95 & 3.9516 $^6$ \\
4778 & A1 CR SR EU & 9217 $^{ 2,3 } $ & 1.414 & 2.18 &2.5616 $^6$ \\
11503 & (A1p Si) & 9429 $^{ 3,4,5 } $ & 1.680 & 2.44 & 1.6093 $^6$ \\
15089 & A4 SR & 8250 $^{ 1 } $ & 1.391 & 2.07 & 1.74 $^7$ \\
15144 & A5 SR CR EU & 8372 $^{ 2,3,4 } $ 1.195 & 1.93 & 15.88 $^6$ \\
24712 & A9 SR EU CR & 7257 $^{ 2,3,4 } $ & 0.885 & 1.61 & 12.4617 $^6$ \\
27309 & A0 SI Cr & 12000 $^{ 1 } $ & 1.975 & 3.04 & 1.10496 $^6$ \\
29305 & A0 SI & 12052 $^{ 2,3,4,5 } $ & 2.303 & 3.47 & 2.9432 $^7$ \\
40312 & A0 SI & 10339 $^{ 1 } $ & 2.411 & 3.50 & 3.61866 $^6$ \\
54118 & A0 SI & 10767 $^{ 2,3,4,5 } $ & 1.872 & 2.78 & 3.27533 $^6$ \\
56022 & A1 SI CR SR & 9819 $^{ 2,3,4 } $ & 1.525 & 2.34 & 0.92 $^8$ \\
62140 & A8 SR EU & 8100 $^{ 1 } $ & 1.123 & 1.85 & 4.28488 $^6$ \\
65339 & A2p... & 8717 $^{ 2,3,5 } $ & 1.475 & 2.18 & 8.0267 $^6$ \\
72968 & A2 SR CR & 9500 $^{ 1 } $ & 1.498 & 2.28 & 4.66548 $^6$ \\
83368 & A8 SR CR EU & 7724 $^{ 2,3 } $ & 1.1474 & 1.83 & 2.851962 $^6$ \\
89822 & A0 HG SI SR & 10000 $^{ 1 } $ & 1.954 & 2.79 & 7.5586 $^6$ \\
90763 & A1sp... & 9038 $^{ 2,3,4 } $ & 1.268 & 2.04 & 3.57 $^7$ \\
96616 & A3 SR & 9472 $^{ 2,3,4 } $ & 1.788 & 2.55 & 2.44 $^7$ \\
107452 & A7 SR &      & 3.772 &   	&   \\
108662 & A0 SRCR EU & 10000 $^{ 1 } $ & 1.730 & 2.53 & 5.0805 $^6$ \\
108945 & A3 SR & 8875 $^{ 1 } $ & 1.703 & 3.42 & 1.92442 $^6$ \\
109026 & B5 HE FBL. & 16008 $^{ 2,4,5 } $ & 3.295 & 6.0 &   \\
112185 & A1 CR EU MN & 9375 $^{ 1 } $ & 2.048 & 2.87 & 5.0887 $^6$ \\
112413 & A0 EU SI CR & 11750 $^{ 1 } $ & 2.0233 & 3.05 & 5.46939 $^6$ \\
115735 & B9 HE FBL.SI & 10474 $^{ 2,3,4,5 } $ & 1.835 & 2.70 & 0.77 $^7$ \\
117025 & A2m & 8432$^{ 2,3 } $ & 1.360 & 2.05 &    \\
118022 & A2 CR EU SR & 9250 $^{ 1 } $ & 1.474 & 2.23 & 3.722084 $^6$ \\
119213 & A3 SR CR & 9000 $^{ 1 } $ & 1.312 & 3.06 & 2.44997 $^6$ \\
120198 & A0 EU CR & 9800 $^{ 1 } $ & 1.567 & 2.37 & 1.3879 $^7$ \\
120709 & B5 HE FBL.P & 16656 $^{ 2,4,5 } $ & 2.668 &  &   \\
124224 & B9 SI & 12500 $^{ 1 } $ & 2.006 & 3.17 & 0.520675 $^6$ \\
125248 & A1 EU CR & 9542 $^{ 3,4 } $ & 1.524 & 2.31 & 9.2954 $^6$ \\
128898 & A9 SR EU & 7822 $^{ 2,3 } $ & 1.047 & 2.77 & 4.4794 $^6$ \\
130559 & A1 SR CR EU & 9607 $^{ 2,3,4 } $ & 1.568 & 2.35 &   \\
133652 & Ap Si & 12837 $^{ 2,3,4,5 } $ & 1.797 & & 2.304 $^6$ \\
134214 & F2 SR EU CR & 7335 $^{ 3,4 } $ & 0.832 & 1.58 & 248 $^7$ \\
137909 & A9 SR EU CR & 7753 $^{ 1 } $ & 1.524 & 2.18 & 18.4868 $^6$ \\
137949 & F0 SR EU CR & 7545 $^{ 1 } $ & 1.125 & 1.80 & 11.13313 $^6$ \\
140160 & A1 SR CR EU & 9250 $^{ 1 } $ & 1.509 & 2.26 & 1.59584 $^6$ \\
140728 & A0 SI Cr & 10083 $^{ 3,4,5 } $ & 1.501 & 2.35 & 1.296 $^7$ \\
148112 & A0 CR EU & 9624$^{ 2,3,4 } $ & 1.868 & 2.62 & 3.04296 $^6$ \\
148898 & A6 SR CR EU & 8327 $^{ 2,3,4 } $ & 1.589 & 2.27 & 1.8 $^7$ \\
151199 & A3 SR & 8616 $^{ 2,3,4 } $ & 1.366 & 2.07 & 6.14 $^7$ \\
152107 & A3 SR CR EU & 8723 $^{ 2,3,4 } $ & 1.469 & 2.18 & 3.86778 $^6$ \\
170000 & A0 SI & 10543 $^{ 3 } $ & 2.260 & 3.27 & 1.71649 $^6$ \\
170397 & A0 SI CR EU & 9921 $^{ 2,3,4,5 } $ & 1.458 & 2.30 & 2.25454 $^6$ \\
176232 & A6 SR CR EU & 8000 $^{ 1 } $ & 1.270 & 1.94 & 6.5 $^7$ \\
201601 & A9 SR EU & 8000 $^{ 1 } $ & 1.112 & 1.83 & 27027 $^6$ \\
202627 & A1 SI CR SR & 8846 $^{ 2,3,4 } $ &  2.17 & 1.45 &    \\
203006 & A2 CR EU SR & 10101 $^{ 2,3,4,5 } $ & 1.599 & 2.43 & 2.122 $^7$ \\
206742 & A0 SI & 10214 $^{ 2,3,4,5 } $ & 1.890 & 2.73 &   \\
220825 & A1 CR SR EU & 9860 $^{ 2,3,4 } $ & 1.405 & 2.25 & 1.14077 $^6$ \\
221760 & A2 SR CR EU & 8581 $^{ 2,3,4 } $ & 1.882 & 2.62 & 12.5 $^7$ \\
223640 & B9 SI SR CR & 13000 $^{ 1 } $ & 2.136 & 3.42 & 3.735239 $^6$ \\

\hline \hline

\end{longtable}


\begin{thebibliography}{}
%
\bibitem{name1}
Abt, H. A., Levato, H., \& Grosso, M., 2002, ApJ 573, 359
\bibitem{name2}
Abt, H. A., \& Morrell, N. I., 1995, ApJS 99, 135
\bibitem{name3}
Adelman, S.J., Pyper, D.M., Shore, S.N., White, R.E., \& Warren, W.H., 1989, A\&AS 81, 221
\bibitem{name4}
Babcock, H. W., 1947, ApJ 105, 105
\bibitem{name5}
Balona, L. A., 1994, MNRAS 268, 119
\bibitem{name6}
Bohlender, D. A., Landstreet, J. D., \& Thompson, Ian B., 1993, A\&A 269, 355
\bibitem{name7}
Brown, A. G. A., \& Verschueren, W., 1997, A\&A 319, 811
\bibitem{name8}
Bychkov, V. D., Bychkova, L. V., \& Madej, J., 2006, MNRAS 365, 585
\bibitem{name9}
Bychkov, V. D., Bychkova, L. V., \& Madej, J., 2005, A\&A 430, 1143
\bibitem{name10}
Bychkov, V. D., Bychkova, L. V., \& Madej, J., 2003, A\&A 407, 631
\bibitem{name11}
Catalano, F. A., \& Renson, P., 1997, A\&AS 127, 421
\bibitem{name12}
Donati, J.-F., Semel, M., Carter, B. D., Rees, D. E., \& Collier Cameron, A., 1997, MNRAS 291, 658
\bibitem{name13}
ESA, 1997, {\it The Hipparcos and Tycho Catalogues}
\bibitem{name15}
Gray, D. F., 2005, {\it The Observations and Analysis of Stellar Photospheres} D. F. Gray (ed). UK, Cambridge University Press, 3rd Ed.
\bibitem{name16}
Hatzes, A. P., 1991, MNRAS 253, 89
\bibitem{name17}
Hauck, B., \& North, P., 1993, A\&A 269, 403
\bibitem{name18}
Hoffleit, D.,\& Jaschek, C., 1991, {\it The Bright Star Catalogue} D. Hoffleit and C. Jascheck (eds). New Haven, Conn., Yale University Observatory,  5th rev.ed.
\bibitem{noemie}
Johnson, N., M.Sc. thesis, Royal Military College of Canada
\bibitem{name19}
Kreidl, T. J., Kurtz, D. W., Schneider, H., van Wyk, F., Roberts, G., Marang, F., \& Birch, P. V., 1994, MNRAS 270, 115
\bibitem{name20}
Kupka, F., \& Bruntt, H., 2001, JAD 7Q, 8
\bibitem{name50}
Landstreet 2006, personal correspondence
\bibitem{name21}
Mermilliod, J.-C., Mermilliod, M., \& Hauck, B., 1997, A\&AS 124, 349
\bibitem{name22}
Renson, P., Gerbaldi, M., \& Catalano, F. A., 1991, A\&AS 90, 91
\bibitem{name23}
Rice, J. B., \& Wehlau, W. H., 1994, A\&A 291, 825
\bibitem{name24}
Romanyuk, I. I., 2000, {\it Magnetic Fields of Chemically Peculiar and Related Stars}, Proceedings of the International Meeting, Special Astrophysical Observatory of Russian AS,  Yu.V. Glagolevskij, I.I. Romanyuk,(eds), p.18-27 
\bibitem{name25}
Royer, F., Grenier, S., Baylac, M.-O., Gomez, A. E., \& Zorec, J., 2002, A\&A 393, 897
\bibitem{name26}
Schaller, G., Schaerer, D., Meynet, G., \& Maeder, A., 1992, A\&AS 996, 296
\bibitem{name27}
Shorlin, S. L. S., Wade, G. A., Donati, J.-F.; Landstreet, J. D., Petit, P., Sigut, T. A. A., \& Strasser, S., 2002, A\&A 392, 637
\bibitem{name28}
Stepien, K., 1994, in {\it Chemically peculiar and magnetic stars}, J. Zverko and J. Ziznovsky (eds). Tantranska Lomnica, Slovak Academy of Sciences, p. 8
\bibitem{name14}
Van Genderen, A. M., 1971, A\&A 14, 48
\bibitem{name29}
Wade, G. A., Donati, J.-F, Landstreet J. D., \& Shorlin, S.L.S., 2000, MNRAS 313, 851
\bibitem{wolff}
Wolff, S.C., 1968, PASP 80, 281


\end{thebibliography}
\end{document}